\documentclass[]{article}
\usepackage{lmodern}
\usepackage{amssymb,amsmath}
\usepackage{ifxetex,ifluatex}
\usepackage{fixltx2e} 
\ifnum 0\ifxetex 1\fi\ifluatex 1\fi=0 
  \usepackage[T1]{fontenc}
  \usepackage[utf8]{inputenc}
\else 
  \ifxetex
    \usepackage{mathspec}
  \else
    \usepackage{fontspec}
  \fi
  \defaultfontfeatures{Ligatures=TeX,Scale=MatchLowercase}
\fi
\IfFileExists{upquote.sty}{\usepackage{upquote}}{}
\IfFileExists{microtype.sty}{%
\usepackage{microtype}
\UseMicrotypeSet[protrusion]{basicmath} 
}{}
\usepackage[margin=1in]{geometry}
\usepackage{hyperref}
\hypersetup{unicode=true,
            pdftitle={Discovering Language of the Stocks},
            pdfauthor={Marko Poženel and Dejan Lavbič},
            pdfborder={0 0 0},
            breaklinks=true}
\usepackage{natbib}
\usepackage{longtable,booktabs}
\usepackage{graphicx,grffile}
\makeatletter
\def\maxwidth{\ifdim\Gin@nat@width>\linewidth\linewidth\else\Gin@nat@width\fi}
\def\maxheight{\ifdim\Gin@nat@height>\textheight\textheight\else\Gin@nat@height\fi}
\makeatother
\setkeys{Gin}{width=\maxwidth,height=\maxheight,keepaspectratio}
\IfFileExists{parskip.sty}{%
\usepackage{parskip}
}{
\setlength{\parindent}{0pt}
\setlength{\parskip}{6pt plus 2pt minus 1pt}
}
\providecommand{\tightlist}{%
  \setlength{\itemsep}{0pt}\setlength{\parskip}{0pt}}
\ifx\paragraph\undefined\else
\let\oldparagraph\paragraph
\renewcommand{\paragraph}[1]{\oldparagraph{#1}\mbox{}}
\fi
\ifx\subparagraph\undefined\else
\let\oldsubparagraph\subparagraph
\renewcommand{\subparagraph}[1]{\oldsubparagraph{#1}\mbox{}}
\fi

\let\rmarkdownfootnote\footnote%
\def\footnote{\protect\rmarkdownfootnote}

\usepackage{titling}

  \title{Discovering Language of the Stocks}
    \pretitle{\vspace{\droptitle}\centering\huge}
  \posttitle{\par}
    \author{Marko Poženel and Dejan Lavbič}
    \preauthor{\centering\large\emph}
  \postauthor{\par}
    \date{}
    \predate{}\postdate{}

\usepackage{booktabs}
\usepackage{longtable}
\usepackage{array}
\usepackage{multirow}
\usepackage{wrapfig}
\usepackage{float}
\usepackage{colortbl}
\usepackage{pdflscape}
\usepackage{tabu}
\usepackage{threeparttable}
\usepackage{threeparttablex}
\usepackage[normalem]{ulem}
\usepackage{makecell}
\usepackage{xcolor}

\begin{document}
\maketitle

\begin{quote}
Marko Poženel and \textbf{Dejan Lavbič}. 2019. \textbf{Discovering
Language of Stocks},
\href{http://ebooks.iospress.nl/ISBN/978-1-61499-941-6}{Frontiers in
Artificial Intelligence and Applications - \textbf{Databases and
Information Systems X}}, 315, pp.~243 - 258, IOS Press.
\end{quote}

\section*{Abstract}\label{abstract}
\addcontentsline{toc}{section}{Abstract}

Stock prediction has always been attractive area for researchers and
investors since the financial gains can be substantial. However, stock
prediction can be a challenging task since stocks are influenced by a
multitude of factors whose influence vary rapidly through time. This
paper proposes a novel approach (Word2Vec) for stock trend prediction
combining NLP and Japanese candlesticks. First, we create a simple
language of Japanese candlesticks from the source OHLC data. Then,
sentences of words are used to train the NLP Word2Vec model where
training data classification also takes into account trading
commissions. Finally, the model is used to predict trading actions. The
proposed approach was compared to three trading models Buy \& Hold, MA
and MACD according to the yield achieved. We first evaluated Word2Vec on
three shares of Apple, Microsoft and Coca-Cola where it outperformed the
comparative models. Next we evaluated Word2Vec on stocks from Russell
Top 50 Index where our Word2Vec method was also very successful in test
phase and only fall behind the Buy \& Hold method in validation phase.
Word2Vec achieved positive results in all scenarios while the average
yields of MA and MACD were still lower compared to Word2Vec.

\section*{Keywords}\label{keywords}
\addcontentsline{toc}{section}{Keywords}

Stock price prediction, Word2Vec, Japanese candlesticks, Trading
strategy, NLP

\section{Introduction}\label{introduction}

Investors and the research community have always found forecasting
trends and the future value of stocks an interesting topic. Moderately
accurate prediction in stock trends can result in high financial
benefits and hedge against market risks \citep{kumar_forecasting_2006}.
Given the attractiveness of the research area, the number of successful
research papers is still quite low. The main reason is usually that
nobody wants to publish an algorithm that solves one of the issues that
might be most profitable. One of the reasons could be the fact that
investors for a long time accepted the Efficient Market Hypothesis (EMH)
\citep{fama_efficient_1960}. Hypothesis states that prices immediately
incorporate all available information about a stock and only new
information is able to change price movements
\citep{cavalcante_computational_2016}, so abnormal yields are not
possible only based on studying the evolution of stock price's past
behavior
\citep{tsinaslanidis_prediction_2014, ballings_evaluating_2015}.

In the last decades, some economists are skeptical about EMH and
sympathetic to the idea that stock prices are partially predictable.
Others claim that stock markets are more efficient and less predictable
than many recent academic papers would have us believe
\citep{malkiel_efficient_2003}. Nonetheless, a lot of approaches to
forecasting the future stock values have been explored and presented
\citep{cavalcante_computational_2016, ballings_evaluating_2015}.

The main goal of stock market analysis is to better understand stock
market in order to be able to take better decisions. Two the most common
approaches for stock market analysis are fundamental
\citep{abad_fundamental_2004} and technical analysis
\citep{taylor_use_1992}. The biggest difference between these two
approaches is in the stock market attributes that are taken into account
in the analysis. Fundamental analysis inspects the basic company
properties such as: the company size, price / profits ratio, assets, and
other financial aspects. Often, the marketing strategy, management
policy, and company innovation are also taken into account. Fundamental
analysis can be improved by including external, political and economic
factors like legislation, market trends and data available on-line
\citep{cavalcante_computational_2016}.

On the other hand, technical analysis is not interested in analyzing
internal and external characteristics of the company. It rather focuses
solely on trading, analyses stock chart patterns and volume of trading,
monitors trading activities, leaving out a number of subjective factors.
Technical analysis is based on the assumption that all internal and
external factors that affect company's stock price are already
indirectly included in the stock price. The tools used by technical
analysis are charting, relative strength index, moving averages, on
balance volumes and others. Technical analysis is based on historical
data to predict future stock trends. With EMH in mind, it could be
inferred that this market analysis approach will not be effective.
However, several scientific papers published in literature using
technical analysis have presented successful approaches in stock
prediction \citep{cavalcante_computational_2016}.

With technical analysis focusing only on stock prices, prediction of
future stock trends can be translated to pattern recognition problem,
where financial data are modelled as time series
\citep{teixeira_method_2010}. As a result, several tools and techniques
are available ranging from traditional statistic modelling to
computational intelligence algorithms
\citep{cavalcante_computational_2016}.

The candlestick trading strategy
\citep{lu_efficient_2011, nison_japanese_1991} is a very popular
technical method to reveal the growth and decline of the demand and
supply in financial markets. It is one of the oldest technical analyses
techniques with origins in \(18^{\text{th}}\) century where it was used
by Munehisa Homma for trading with rice. He analysed rice prices back in
time and acquired huge insights to the rice trading characteristics.
Japanese candlestick charting technique is a primary tool to visualize
the changes in a commodity price in a certain time span. Almost every
software and on-line charting packages \citep{jasemi_modern_2011}
available today include candlestick charting technique. Although the
researchers are not in complete agreement about its efficiency, many
researchers are investigating its potential use in various fields
\citep{do_prado_effectiveness_2013, jasemi_modern_2011, lu_tests_2012, kamo_hybrid_2009, lu_profitability_2014}.
To visualize Japanese candlestick at a certain time grain (e.g.~day,
hour), four key data components of a price are required: starting price,
highest price, lowest price and closing price. This tuple is called OHLC
(Open, High, Low, Close). When the candlestick body is filled, the
closing price of the session was lower than the opening price. If the
body is empty, the closing price was higher than the opening price. The
thin lines above and below the rectangle body are called shadows and
represent session's price extremes. There are many types of Japanese
candlesticks with their distinctive names. Each candlestick holds
information on trading session and becomes even more important, when it
is an integral part of certain sequence.

The goal of our research is defining a simplified language of Japanese
candlesticks from OHLC data. This simplified OHLC language is than used
as an input for Word2Vec algorithm \citep{mikolov_distributed_2013} that
can learn the vector representations of words in the high-dimensional
vector space. We believe that it is possible to learn rules and patterns
using Word2Vec and use this knowledge to predict future trends in stock
value. Despite many developed models and predictive techniques,
measuring performance of the stock prediction models can present a
challenge. For example, Jasemi et al. \citep{jasemi_modern_2011} used
hit ratio to evaluate the performance of the models but neglected
financial success of a model. Therefore, one of the research goals of
this paper is also to utilize a simple method for testing the
performance of forecasting models, the result of which is the financial
success or yield of the tested model.

The remaining paper is organized as follows. Section \ref{related-work}
contains a literature overview. Section \ref{proposed-forecasting-model}
is dedicated to a detailed overview of the proposed forecasting model.
In Section \ref{evaluation} model evaluation and performance metrics are
presented. Section \ref{conclusion-and-future-work} presents the
conclusions and future work.

\section{Related work}\label{related-work}

Stock forecasting is one of the major activities of financial firms in
order to make investment decisions. What is more, stock forecasting can
be considered as one of the main challenges of time-series and machine
learning science community \citep{tay_application_2001}. However, the
stock price prediction is very difficult task since many parameters have
to be considered, where many of them can not be easily modelled.
Financial markets are complex, dynamic, non-linear systems influenced by
political events, economic outlook, traders' expectations
\citep{huang_forecasting_2005}. One of the main problems with predicting
stock price direction is also the huge amount of data. The data sets are
too big to handle with more traditional methods
\citep{ballings_evaluating_2015}. In literature, several approaches for
stock price prediction have been proposed in recent years
\citep{teixeira_method_2010, cavalcante_computational_2016}. One of the
best performing algorithms for stock prediction appears to be Support
Vector Machines (SVM), while one of the most common computing techniques
used for forecasting financial time series are (Artificial) Neural
Networks (A)NN \citep{ballings_evaluating_2015}.

The main reason NN have become very popular for financial forecasting is
because this computing technique is able to handle data that are
non-linear, contain discontinuities and high-frequency polynomial
components \citep{liu_fluctuation_2012}. NN are data-driven and
self-adaptive methods able to capture non-linear behaviours of time
series without any statistical assumptions about the data
\citep{lu_financial_2009}.

\citet{martinez_artificial_2009} proposed a day-trading system based on
a ANN that forecasts daily minimum and maximum stock prices. Presented
approach uses a multi-layer feed-forward neural network trained by the
back-propagation algorithm. The NN uses three classical layers: input,
output and one hidden layer. Multi-layer NN is used to learn the
relationship between variables and to predicts prices. A set of trading
rules is used to signalize the investor the best time to buy or sell
stocks.

\citet{tay_application_2001} proposed a NN technique, support vector
machine (SVM) to forecast financial time series. It was compared to the
multi-layer back-propagation (BP) neural network and achieved
significantly better results. Source for experimental data was Chicago
Mercantile Market that contains data for S\&P 500, US 30-year government
bonds, German 10-year government bonds and others. SVMs provide a
promising alternative to the BP neural network for financial time series
forecasting. SVMs forecast significantly better than the BP in the
majority of stocks futures and slightly better in the German 10-year
government bonds.

\citet{lu_efficient_2011} proposed an efficient cerebellar model
articulation controller neural network (CMAC NN) scheme for forecasting
stock prices. A CMAC is a supervised NN using a least mean square
algorithm in training phase. The proposed method improves traditional
CMAC NN. It employs a fast hash coding to speed up the many-to-few
mappings and reduces generalization error by using a high quantization
resolution. They compared the performance of CMAC NN with support vector
regression (SVR) and a back-propagation NN (BPNN) and achieved superior
results. The proposed scheme is easier to use than traditional
statistical and spectral analytical methods.

\citet{liu_fluctuation_2012} forecast the price fluctuation by an
improved Legendre NN. They use time-variant data training set, where
older data affect prediction values differently as recent one. A
tendency function and Random Brownian volatility function is used to
define the weight of the time-series data. Experimental results show
that proposed approach adapts the volatile market movements
outperforming simple Legendre NN.

In \citep{wang_forecasting_2015} Wang et al. introduced the stochastic
time effective neural network (STNN) with principal component analysis
model (PCA) to forecast stock indexes SSE, HS300, S\&P500, and DJIA. In
the training phase, PCA approach is used to obtain principal components
from the source data. The financial price series prediction is performed
by STNN model. The Brownian motion is used to define the degree of
impact of historical data. The proposed NN outperforms the traditional
back-propagation neural network (BPNN), PCA-BPNN and STNN in financial
time series forecasting.

\citet{hafezi_bat-neural_2015} also included NN in their stock price
prediction model called bat-neural network multi-agent system (BNNMAS).
They proposed a four layer BNNMAS architecture where each layer has its
goals and sub-goals coordinated with other layers to increase prediction
accuracy. They tested the model on DAX stock data in eight years time
span, which included financial crisis in 2008. Compared to fundamental
and technical analysis, BNNMAS achieved good accuracy of the model in a
long term periods.

In literature, the authors use the predictive power of Japanese
candlesticks mostly on the basis of expert knowledge and rules that are
based on past patterns. \citet{lu_tests_2012} used the four-digit
numbers approach to categorize two-day candlestick patterns and tested
the approach on Taiwanese stock market. They demonstrated that
candlestick analysis has value for investors, what violates efficient
markets hypothesis \citep{fama_efficient_1960}. They found some existing
patterns not profitable, and showing two new patterns as profitable.

\citet{kamo_hybrid_2009} implemented a study that illustrates the basic
candlestick patterns and the standard IF-THEN fuzzy logic model. They
employed generalized regression neural networks (GRNN) with rule-based
fuzzy gating network. Every GRNN handles one OHLC attribute value, which
are then combined to the final prediction with fuzzy logic model. They
compared the approach to candlestick method based on GRNN with a simple
gating network and it performed better.

\citet{jasemi_modern_2011} also used neural networks (NN) for technical
analysis of Japanese candlesticks. In their approach NN is not used just
to learn the candlestick lines and create a set of static rules, but
rather NN continuously analyses input data and updates technical rules.
They focused on discovering turning points in prices to trigger buying
and selling actions at the best time. The presented approach yields
better results than approach using static selection of rules and input
signals. Unfortunately, the authors do not present the data, whether the
financial success is obtained in the stock market.

\citet{martiny_unsupervised_2012} presented the method that utilizes
unsupervised machine-learning for automatically discovering significant
candlestick patterns from a time series of price data. OHLC data is
first clustered using Hierarchical Clustering, then a Naive Bayesian
classifier is used to predict future prices based on daily sequences.
The performance of the proposed technique was measured by the percentage
of properly triggered sell/buy signals. Although authors in
\citep{keogh_clustering_2005} argue that clustering of time-series
subsequences is meaningless.

\citet{savic_tvorba_2016} explored the idea of combining Japanese
Candlestick language with Natural Language Processing algorithm to
implement a basic stock value trend forecasting algorithm. The idea was
tested on a sample stock data, where the method achieved promising
results. Our work is inspired by the results achieved by Savić.

In this work we present a novel method for forecasting future stock
value trends that combines technical analysis method of Japanese
candlesticks with deep learning. The proposed model integrates Word2Vec,
which is commonly used for the processing of unstructured texts into
technical analysis. Word2Vec can find the deep semantic relationships
between words in the document. In their work, \citet{zhang_chinese_2015}
confirmed that Word2Vec is suitable for Chinese texts clustering and
they also state that Word2Vec shows superior performance in texts
classification and clustering in English
\citep{mikolov_distributed_2013, mikolov_efficient_2013, mikolov_linguistic_2013}.
We have employed the Word2Vec approach in the stock value trend
prediction and to the best of our knowledge, none of the existing
researches uses Word2Vec for forecasting future stock value trends.

\section{Proposed forecasting model}\label{proposed-forecasting-model}

A combination of various machine learning methods in a novel and
innovative way was combined in the proposed forecasting model. The basic
assumption behind the proposed approach is that Japanese candlesticks
are not only powerful tool for visualizing OHLC data, but also contain
predictive power
\citep{jasemi_modern_2011, lu_tests_2012, kamo_hybrid_2009, lu_profitability_2014}.

Various sequences of Japanese candlesticks are used to forecast the
value of a stock in our approach. The foundation for stocks' language
(i.e.~words) is defined by Japanese candlesticks. A language in general
consists of words and patterns of words that can be further grouped into
sentences that express some deeper meaning. The proposed model relies on
the similarities with the natural language.

\begin{figure}

{\centering \includegraphics[width=0.63\linewidth]{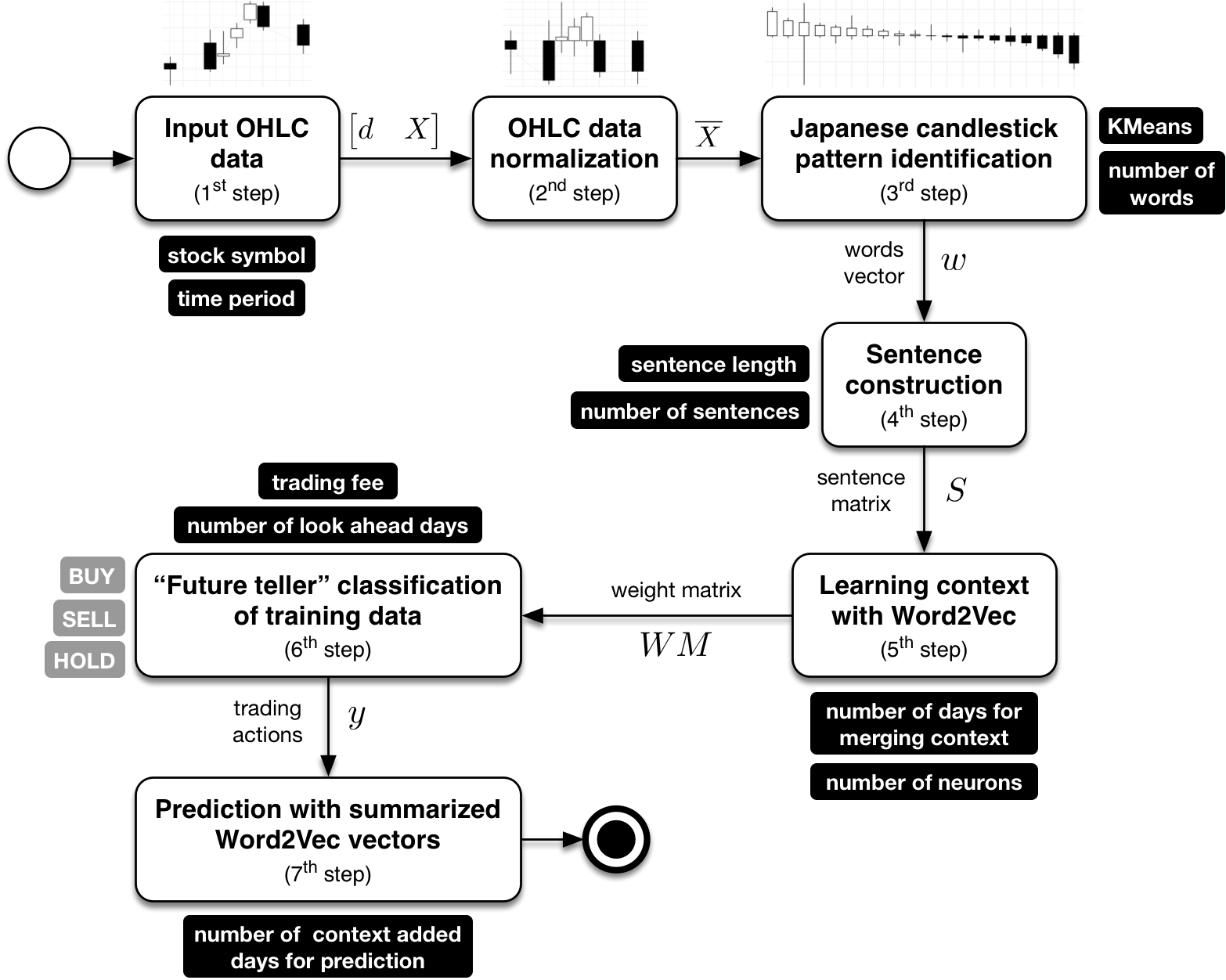}

}

\caption{Steps of proposed forecasting model}\label{fig:Model}
\end{figure}

In the beginning of the forecasting process, the transformation of OHLC
data is performed and results in a simplified language of Japanaese
candlesticks, i.e.~stocks' language. The acquired language is then
processed with the NLP algorithm Word2Vec
\citep{mikolov_distributed_2013} where we train the model with given
characteristics and the legality of the proposed stocks' language. For
predicting future trends in a stock value the trained model is then
employed. The approach is depicted in Figure \ref{fig:Model}, with
detailed description provided in the following subsections.

\subsection{OHLC data}\label{ohlc-data}

For a given stock we observe the \emph{\textbf{input data}} on a trading
day basis for \(\boldsymbol{n_d}\) \emph{\textbf{trading days}} as
defined in the following matrix

\begin{equation}
\begin{bmatrix}
d_{(\text{$1$ $\times$ $n_d$})} & X_{(\text{$4$ $\times$ $n_d$})}
\end{bmatrix}
=
\begin{bmatrix}
\begin{array}{cc}
d_1 \\
d_2 \\
\dots \\
d_{n_d}
\end{array}
\left|
\begin{array}{*{4}c}
O_1 & H_1 & L_1 & C_1 \\
O_2 & H_2 & L_2 & C_2 \\
\dots & \dots & \dots & \dots \\
O_{n_d} & H_{n_d} & L_{n_d} & C_{n_d}
\end{array}
\right.
\end{bmatrix}
\label{eq:dX}
\end{equation}

where \(d_{(\text{$1$ $\times$ $n_d$})}\) is a vector of trading days
and \(X_{(\text{$4$ $\times$ $n_d$})}\) is a matrix of OHLC trading
data.

OHLC tuples are Japanese candlesticks presentation with individual four
attributes that denote absolute value in time. Raw OHLC data in Equation
\eqref{eq:dX} are convenient for graphical presentation but are not most
suitable for further processing.

\subsection{Data normalization}\label{OHLC-data-normalization}

We are interested in the shape of Japanese candlesticks and not an
absolute value, so the OHLC tuples were normalized by dividing OHLC data
attributes (Open, High, Low, Close) with Open attribute as follows

\begin{equation}
norm\big(\langle O,H,L,C \rangle\big) = \langle 1, \frac{H}{O}, \frac{L}{O}, \frac{C}{O}\rangle: X \to \overline{X}
\label{eq:norm}
\end{equation}

The employment of the transformation from Equation \eqref{eq:norm} results
in a new input trading data matrix

\begin{equation}
\overline{X}_{(\text{$4$ $\times$ $n_d$})} =
\begin{bmatrix}
1 & \frac{H_1}{O_1} & \frac{L_1}{O_1} & \frac{C_1}{O_1} \\\
1 & \frac{H_2}{O_2} & \frac{L_2}{O_2} & \frac{C_2}{O_2} \\\
\dots & \dots & \dots & \dots \\\
1 & \frac{H_{n_d}}{O_{n_d}} & \frac{L_{n_d}}{O_{n_d}} & \frac{C_{n_d}}{O_{n_d}}
\end{bmatrix}
\label{eq:X-norm}
\end{equation}

where the shape of Japanese candlesticks is retained.

\subsection{Word Pattern
Identification}\label{word-pattern-identification}

The majority of forecasting models employing Japanese candlesticks have
a drawback of using predefined shapes of candlesticks
\citep{martiny_unsupervised_2012}.

Our approach uses automatic detection of candlestick clusters with
unsupervised machine learning methods that were beneficial in previous
research \citep{martiny_unsupervised_2012, jasemi_modern_2011}.

The reason behind using K-Means clustering was to limit the number of
possible OHLC shapes (i.e.~words of stocks' language) while still being
able to influence the unsupervised training process by defining selected
threshold for maximum number of different words.

In the process we define the \emph{\textbf{maximum number of words}} in
stocks' language as \(\boldsymbol{n_w}\) and employ
\emph{\textbf{K-Means clustering}} algorithm to transform input data
\(\overline{X}\) to vector \(w\) as follows

\begin{equation}
KMeans\big(n_w\big): \overline{X} \to w
\label{eq:KMeans}
\end{equation}

where a word \(w_i\) is defined by an individual trading day
\(\overline{X_i}\) and is a representation of a specific Japanese
candlestick (the mean value of cluster \(i\)). The result of KMeans
clustering is a vector

\begin{equation}
w_{(\text{$1$ $\times$ $n_d$})} =
\begin{bmatrix}
w_1 & w_2 & \dots & w_{n_d}
\end{bmatrix}^T
\label{eq:w}
\end{equation}

where given word \(w_i\) is an element from a set of all possible
Japanese candlesticks, where \(i = \big[1,n_w\big]\).

\begin{figure}

{\centering \includegraphics[width=0.63\linewidth]{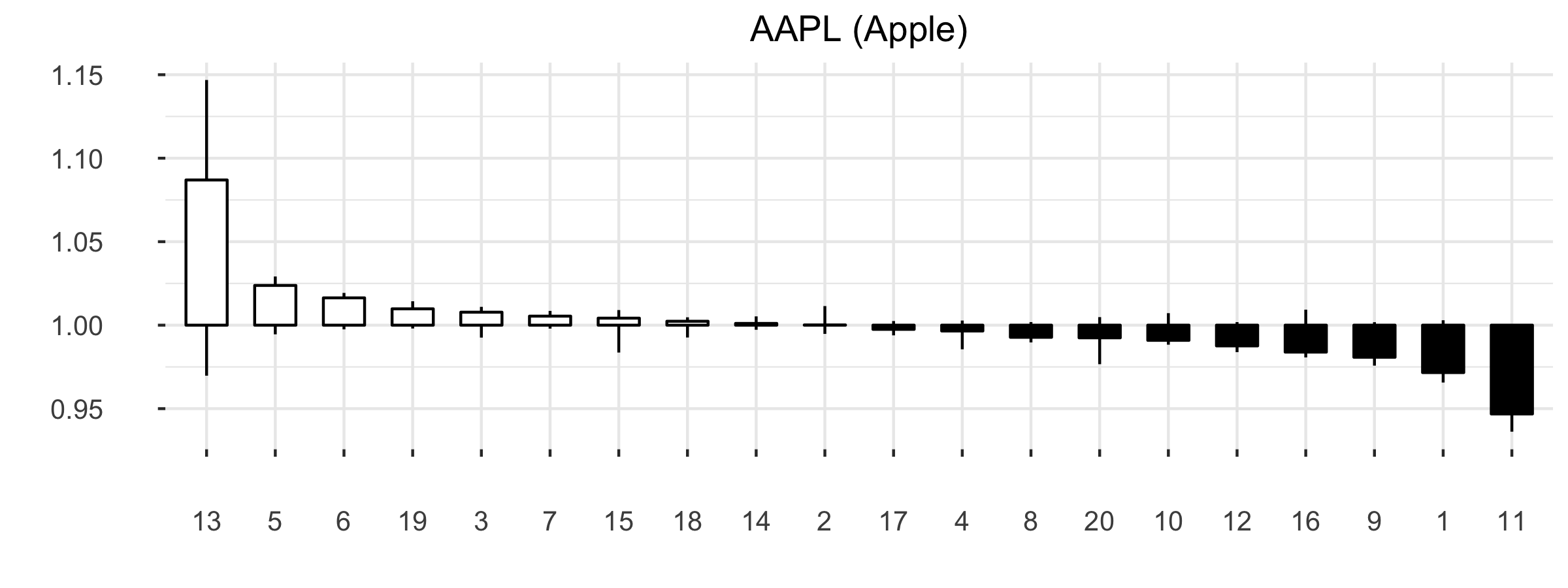}

}

\caption{Example of $20$ OHLC pattern clusters for stock AAPL}\label{fig:Clusters}
\end{figure}

An example of a clustering process for a stock KO (Coca-Cola) is
depicted in Figure \ref{fig:Clusters}, where \(n_w = 20\) was used for
maximum number of words. The value of parameter \(n_w\) is based on the
Silhouette measure \citep{rousseeuw_silhouettes:_1987}, which shows how
well an object lies in within a certain cluster (cohesion) compared to
other clusters (separation). The Silhouette ranges from -1 to +1, where
higher value of average Silhouettes means higher clustering validity. In
defining stocks' language, our aim was also to retain the similarity of
words that also exists in natural language by controlling \(n_w\) and
the Silhouette measure.

\subsection{From Words to Sentences}\label{from-words-to-sentences}

With numerous OHLC tuples the potential set of words for the stocks'
language is virtually infinite. In the previous section we have limited
this to \(n_w\), which directly influences the performance of the
proposed predictive model.

Looking at the analysis of the past movements in the value of stock we
can see that Japanese candlesticks' sequences contain a certain
predictive power \citep{nison_japanese_1991, lu_tests_2012}. Therefore,
we considered past sequences of OHLC as a basis for the stock trend
prediction by forming possible sentences in the future.

The rules for forecasting purposes in proposed model are not predefined
and are rather constructed from sequences of patterns that are acquired
from past movements in stock value.

We specify a \emph{\textbf{sentence length}} \(\boldsymbol{l_s}\) that
defines the number of consecutive words (i.e.~trading days) grouped into
sentences. The \emph{\textbf{number of sentences}} \(\boldsymbol{n_s}\)
is therefore dependent on the number of trading days \(n_d\) and the
sentence length \(l_s\) and is defined as follows

\begin{equation}
n_s = n_d - (l_s - 1)
\label{eq:ns}
\end{equation}

The result of the sentence construction process is a
\emph{\textbf{sentence matrix}} \(\boldsymbol{S}\) of rolling windows of
trading data (more specifically words in stocks' language from vector
\(w\)) from a transformation \(w \to S\). Sentence matrix \(S\) with
\(l_s\) columns (sentence length) and \(n_s\) rows (number of sentences)
is further defined as

\begin{equation}
S_{(\text{$l_s$ $\times$ $n_s$})} =
\begin{bmatrix}
w_1^{'} & w_2^{'} & \dots & w_{l_s}^{'} \\\
w_2^{'} & w_3^{'} & \dots & w_{l_s + 1}^{'} \\\
\dots & \dots & \dots & \dots \\\
w_{n_s}^{'} & w_{n_s + 1}^{'} & \dots & w_{n_d}^{'}
\end{bmatrix}
\label{eq:S}
\end{equation}

It seems that this kind of OHLC language is very simple. However,
considering the number of possible values for each word \(w_i\), a set
of different possible sentences or patterns is enormous. Therefore, the
defined language has a high expressive power and is suitable for
predictive purposes.

\subsection{Word2Vec Training}\label{word2vec-training}

Based on the patterns in OHLC sentences, the model builds the language
context that is then used to perform predictions in the following steps.
The system employs historical data, recognizes existing patterns in
sentences, learns the context of the words and also renews the context
according to new acquired data by employing \emph{\textbf{Word2Vec
algorithm}} \citep{mikolov_distributed_2013} for training the context.

Word2Vec algorithm with skip-gram
\citep{mikolov_distributed_2013, mikolov_efficient_2013} uses a model to
represent words with vectors from large amounts of unstructured text
data. In the training process, Word2Vec acquires vectors for words that
explicitly contain various linguistic rules and patterns by employment
of neural network that contains only one hidden level, so it is
relatively simple. Many of these patterns can be represented as linear
translations. The Word2Vec algorithm has proved to be an excellent tool
for analysing the natural language, for example, the calculation

\[vector(\text{'Madrid'}) - vector(\text{'Spain'}) + vector(\text{'Paris'})\]

yields the result that is closer to the \(vector(\text{'France'})\) than
any other word vector
\citep{mikolov_efficient_2013, mikolov_linguistic_2013}.

For learning context in financial trading with Word2Vec we define
\emph{\textbf{the number of days for merging context}}
\(\boldsymbol{n_{ww}}\) and \textbf{\textit{the number of neurons}}
\(\boldsymbol{n_v}\) in hidden layer weight matrix. Word2Vec algorithm
performs the following transformation

\begin{equation}
W2V\big(S, n_{ww}, n_v\big) : S \to WM
\label{eq:W2V}
\end{equation}

where the result of Word2Vec learning phase is \emph{\textbf{a Weight
Matrix}} \(\boldsymbol{WM}\) with \(n_v\) columns (number of vectors)
and \(n_w\) rows (number of words in stocks' language) and is defined as
follows

\begin{equation}
WM_{(\text{$n_v$ $\times$ $n_w$})} =
\begin{bmatrix}
v_{1,1} & v_{1,2} & \dots & v_{1,n_v} \\\
v_{2,1} & v_{2,2} & \dots & v_{2,n_v} \\\
\dots & \dots & \dots & \dots \\\
v_{n_w,1} & v_{n_w,2} & \dots & v_{n_w,n_v}
\end{bmatrix}
\label{eq:WM}
\end{equation}

with \(v_{i,j}\) as the \(j\)-th vector (weight) of word \(w_i\).

\subsection{Training Data Classification}\label{future-teller}

The proposed model is already capable of using the context that it
learned from historical data for creating OHLC predictions. However, our
aim is that the predictive model would, based on input OHLC sequence,
trigger one of the following actions:

\begin{itemize}
\tightlist
\item
  \(\text{BUY}\),
\item
  \(\text{SELL}\),
\item
  \(\text{HOLD}\) or do nothing.
\end{itemize}

For prediction of the future stock price we label trading days from
matrix \(X\) in training set with \emph{\textbf{trading actions}}
\(\boldsymbol{y}\) where

\begin{equation}
y_{(\text{$1$ $\times$ $n_d$})} =
\begin{bmatrix}
A_1 & A_2 & \dots & A_{n_d}
\end{bmatrix}^T
\label{eq:y}
\end{equation}

and we classify the individual trading day \(y_i\) as \(\text{BUY}\),
\(\text{SELL}\) or \(\text{HOLD}\) based on the number of
\emph{\textbf{look ahead days}} \(\boldsymbol{n_{la}}\) and
\emph{\textbf{the trading fee}} \(\boldsymbol{v_{fee}}\) as follows

\begin{equation}
y_i =
\begin{cases}
  0 : \text{BUY} & \text{$n_{max} \cdot C_j > n_{max} \cdot C_i + 2 \cdot v_{fee}$, $j \in \big[i, i+n_{la}\big]$} \\
  1 : \text{SELL} & \text{$n_{max} \cdot C_j < n_{max} \cdot C_i - 2 \cdot v_{fee}$, $j \in \big[i, i+n_{la}\big]$} \\
  2 : \text{HOLD} & \text{otherwise}
\end{cases}
\label{eq:y-i}
\end{equation}

where \(C_i\) is the stock's close price of a given trading day \(i\)
and \(\boldsymbol{n_{max}} = \big\lceil\frac{e}{C}\big\rceil\) is
\emph{\textbf{the maximum number of stocks to trade}} with
\(\boldsymbol{e}\) as \emph{\textbf{the initial equity}}.

\subsection{Performing Prediction}\label{performing-prediction}

Our proposed model includes classification using the
\emph{\textbf{SoftMax algorithm}} in our Word2Vec neural network (NN).
SoftMax regression is a multinomial logistic regression and it is a
generalization of logistic regression (see Equation \eqref{eq:softmax}).
It is used to model categorical dependent variables (e.g.
\(\text{$0$ : BUY}\), \(\text{$1$ : SELL}\) and \(\text{$2$ : HOLD}\))
and the categories must not have any order (or rank).

The output neurons of Word2Vec NN use Softmax, i.e.~output layer is a
Softmax regression classifier. Based on input sequence, SoftMax neurons
will output probability distribution (floating point values between
\(0\) and \(1\)), and the sum of all these
\(V = \{ \text{BUY, SELL, HOLD}\}\) output values (\(y_k\)) will add up
to \(1\).

\begin{equation}
y_k = P(y = k \mid x) = \frac{e^{x^T w_k}}{\sum_{i = 1}^{n_w}{e^{x^T w_i}}}
\label{eq:softmax}
\end{equation}

Excessive increase of the model parameters due to over-fitting of data
can also affect the model performance. To minimize the aforementioned
problem, we employed least squares regularization that uses cost
function which pushes the coefficients of model parameters to zero and
hence reduce cost function.

For learning any model we have to omit training days without class
prediction, due to look ahead of ``Future Teller'' from section
\ref{future-teller}, where \emph{\textbf{the corrected number of trading
days}} is \(\boldsymbol{\overline{n_d}} = n_d - n_{la}\).

\subsubsection{Basic prediction}\label{basic-prediction}

When building a basic prediction, we use normalized OHLC data from
matrix \(\overline{X}\) (see section \ref{OHLC-data-normalization}) and
vector of trading actions \(y\) from ``Future Teller'' classification
(see section \ref{future-teller}), where SoftMax classifier defines the
following transformation

\begin{equation}
\begin{bmatrix}
\overline{X}_{(\text{$3$ $\times$ $\overline{n_d}$})} & y_{(\text{$1$ $\times$ $\overline{n_d}$})}
\end{bmatrix} =
\begin{bmatrix}
\begin{array}{*{4}c}
\frac{H_1}{O_1} & \frac{L_1}{O_1} & \frac{C_1}{O_1} \\\
\frac{H_2}{O_2} & \frac{L_2}{O_2} & \frac{C_2}{O_2} \\\
\dots & \dots & \dots \\\
\frac{H_{\overline{n_d}}}{O_{\overline{n_d}}} & \frac{L_{\overline{n_d}}}{O_{\overline{n_d}}} & \frac{C_{\overline{n_d}}}{O_{\overline{n_d}}}
\end{array}
\left|
\begin{array}{cc}
A_1 \\\
A_2 \\\
\dots \\\
A_{\overline{n_d}}
\end{array}
\right.
\end{bmatrix}
\to y = f\big(\frac{H}{O}, \frac{L}{O}, \frac{C}{O}\big)
\label{eq:X-y}
\end{equation}

With basic prediction we did not include the context of OHLC candlestics
appearance, which influence the price movement and therefore, the
prediction did not perform well. In the following step prediction with
Word2Vec was performed and taking into account the context by adding
previous days OHLC candlesticks.

\subsubsection{Word2Vec Prediction with
Summarization}\label{word2vec-prediction-with-summarization}

From vector of words \(w\) (see Equation \eqref{eq:w}) and vector of
trading actions \(y\) (see Equation \eqref{eq:y}) in the following format

\begin{equation}
\begin{bmatrix}
w_{(\text{$1$ $\times$ $\overline{n_d}$})} & y_{(\text{$1$ $\times$ $\overline{n_d}$})}
\end{bmatrix} =
\begin{bmatrix}
\begin{array}{cc}
w_1 \\\
w_2 \\\
\dots \\\
w_{\overline{n_d}}
\end{array}
\left|
\begin{array}{cc}
A_1 \\\
A_2 \\\
\dots \\\
A_{\overline{n_d}}
\end{array}
\right.
\end{bmatrix}
\label{eq:wy}
\end{equation}

we replace words \(w_i\) with a Word2Vec representation with \(n_v\)
features vector (hyper parameter) from Weight Matrix
\(WM_{(\text{$n_v$ $\times$ $n_w$})}\) (see Equation \eqref{eq:WM}), where
\(w_i = \big[v_{i,1}, v_{i,2}, \dots, v_{i,n_v}\big]\). Training data in
a matrix \(X_{(\text{$n_v$ $\times$ $\overline{n_d}$})}^{'}\) is defined
as follows

\begin{equation}
\begin{bmatrix}
X_{(\text{$n_v$ $\times$ $\overline{n_d}$})}^{'} & y_{(\text{$1$ $\times$ $\overline{n_d}$})}
\end{bmatrix}
=
\begin{bmatrix}
\begin{array}{*{4}c}
v_{1,1} & v_{1,2} & \dots & v_{1,n_v} \\\
v_{2,1} & v_{2,2} & \dots & v_{2,n_v} \\\
\dots & \dots & \dots & \dots \\\
v_{w_{\overline{n_d}},1} & v_{w_{\overline{n_d}},2} & \dots & v_{w_{\overline{n_d}},n_v}
\end{array}
\left|
\begin{array}{cc}
A_1 \\\
A_2 \\\
\dots \\\
A_{\overline{n_d}}
\end{array}
\right.
\end{bmatrix}
\label{eq:X1-y}
\end{equation}

We add context by \emph{\textbf{adding previous}} \(\boldsymbol{n_m}\)
\emph{\textbf{trading days}} to the current trading day and define a new
input matrix \(X_{(\text{$n_v$ $\times$ $\overline{n_d}^{'}$})}^{''}\),
where \(\overline{n_d}^{'} = \overline{n_d} - n_m\).

Let
\(\boldsymbol{cv_j} = [cv_{1,j}, cv_{2,j}, \dots, cv_{n_v,j}] \in X^{''}\)
be \emph{\textbf{a context vector}} for a given trading day \(j\) (row
\(j\) in matrix \(X^{''}\)), where \(j \in [1, \overline{n_d}^{'}]\) and
\emph{\textbf{contextualized input matrix}} \(\boldsymbol{X^{''}}\) is
defined as follows

\begin{equation}
\begin{bmatrix}
X_{(\text{$n_v$ $\times$ $\overline{n_d}^{'}$})}^{''} & y_{(\text{$1$ $\times$ $\overline{n_d}^{'}$})}
\end{bmatrix} =
\begin{bmatrix}
\begin{array}{*{4}c}
cv_{1,1} & cv_{1,2} & \dots & cv_{1,n_v} \\\
cv_{2,1} & cv_{2,2} & \dots & cv_{2,n_v} \\\
\dots & \dots & \dots & \dots \\\
cv_{w_{\overline{n_d}}^{'},1} & cv_{w_{\overline{n_d}}^{'},2} & \dots & cv_{w_{\overline{n_d}}^{'},n_v}
\end{array}
\left|
\begin{array}{cc}
A_1 \\\
A_2 \\\
\dots \\\
A_{\overline{n_d}^{'}}
\end{array}
\right.
\end{bmatrix}
\label{eq:X2-y}
\end{equation}

where context vector \(cv_j\) is a sum of vectors of \(n_m\) previous
trading days as follows

\begin{equation}
cv_j = \sum_{k = j}^{j + n_m} v_k
\label{eq:cv-j}
\end{equation}

where \(v_k = [v_{1,k}, v_{2,k}, \dots, v_{\overline{n_d},k}]\) is the
\(\text{$k$-th}\) row in matrix \(X^{'}\).

\section{Evaluation}\label{evaluation}

To summarize the findings of the results for Apple (AAPL), Microsoft
(MSFT) and Coca-Cola (KO) shares, the proposed model yielded promising
results. In the test phase, the proposed forecast model combined with
the proposed trading strategy outperformed all comparative models as
depicted in Table \ref{tab:avg-yield-stocks-test-phase}.

\begin{table}[H]

\caption{\label{tab:avg-yield-stocks-test-phase}Average yields of forecasting models on selected stocks at an initial investment of \$10,000 in the test phase}
\centering
\begin{tabular}{lrrrr}
\toprule
 & Buy \& Hold & MA(50,100) & MACD & \textbf{W2V}\\
\midrule
Apple (AAPL) & \$102,557.08 & \$34,915.34 & \$46,452.72 & \textbf{\$182,938.35}\\
Microsoft (MSFT) & -\$2,927.03 & -\$4,140.42 & -\$3,261.15 & \textbf{\$11,109.06}\\
Coca-Cola (KO) & \$1,996.82 & \$2,066.74 & -\$1,108.05 & \textbf{\$4,360.76}\\
\textbf{Average} & \$33,875.62 & \$10,947.21 & \$14,027.84 & \textbf{\$66,136.05}\\
\bottomrule
\end{tabular}
\end{table}

In the validation phase, the performance was a bit lower but the average
yield of the proposed model was still higher than the comparable models
as depicted in Table \ref{tab:avg-yield-stocks-validation-phase}.
However, drawing conclusions based only on three sample shares may not
be meaningful, so we carried out extensive testing on a larger data set
and run confirmatory data analysis.

\begin{table}[H]

\caption{\label{tab:avg-yield-stocks-validation-phase}Average yields of forecasting models on selected stocks at an initial investment of \$10,000 in the validation phase}
\centering
\begin{tabular}{lrrrr}
\toprule
 & Buy \& Hold & MA(50,100) & MACD & \textbf{W2V}\\
\midrule
Apple (AAPL) & \$28,611.11 & \$32,339.63 & \$6,619.31 & \textbf{\$57,543.47}\\
Microsoft (MSFT) & \textbf{\$20,316.42} & \$1,809.31 & \$2,477.12 & \$10,603.90\\
Coca-Cola (KO) & \textbf{\$5,547.81} & \$3,583.26 & -\$4,220.57 & \$3,163.32\\
\textbf{Average} & \$18,158.45 & \$12,577.40 & \$1,625.28 & \textbf{\$23,770.19}\\
\bottomrule
\end{tabular}
\end{table}

For the final test set we selected stocks from Russell Top 50 Index,
which includes 50 stocks of the largest companies (market cap and
current index membership) in the U.S stock market. The forecasting model
was tested for each stock separately. Thus, for each of the 50 stocks,
the prediction model was trained based on past stock values of the
particular stock. In the test phase, the model parameters were adjusted
that the model achieved highest yield for particular stock. The trained
model with parameters tuned for the particular stock was then tested on
validation set. Table \ref{tab:avg-yield-russel-10k} shows average yield
achieved by the proposed Word2Vec (W2V) model as well as yield achieved
by comparative models (Buy \& Hold, Moving Average and MACD) for the
test and validation phase. In the test phase, average yield of the
proposed W2V model was much higher than yield of the comparative models.
However, in the validation phase the results were not as good as in the
test phase. The average yield of Moving Average and MACD models were
still smaller, while Buy \& Hold outperformed our model.

\begin{table}[H]

\caption{\label{tab:avg-yield-russel-10k}Average yields of forecasting models on stocks of the Russell Top 50 index at an initial investment of \$10,000}
\centering
\begin{tabular}{lrrrr}
\toprule
 & Buy \& Hold & MA(50,100) & MACD & \textbf{W2V}\\
\midrule
Russell Top 50 Index - \textbf{Test phase} & \$2,818.98 & \$1,073.06 & -\$482.04 & \textbf{\$11,725.25}\\
Russell Top 50 Index - \textbf{Validation phase} & \textbf{\$16,590.83} & \$6,238.43 & \$395.10 & \$10,324.24\\
\bottomrule
\end{tabular}
\end{table}

A more detailed results for individual stocks at the test phase is
presented in Table \ref{tab:avg-yield-russel-10k-individual}. In the
test phase, our model generates profit for all except one stock
(i.e.~JNJ), where zero profit is achieved. What is more, our model
outperformed the comparative models in all but three cases (stocks SLB,
DIS and JNJ). In the validation phase, the results are worse but still
encouraging. Only in \(14\%\) of cases the model gave negative yield,
while in \(16\%\) cases the model outperformed all comparative models.
In \(30\%\) of cases the model was the second best model. What is more,
in \(7\) cases the model's yield was very close to the yield of the best
method.

Average yield gives us some information about the model's performance.
However, based on average yield we are unable to conclude whether the
proposed model yields statistically significant better results than
comparative models. To get statistically significant results, we carried
out statistical tests. We have two nominal variables: forecasting model
(e.g.~Buy \& Hold vs.~W2V) and individual stock (e.g.~IBM, AAPL, MSFT,
GOOGL etc.) and one measurement variable (stock yield). We have two
samples in which observations in one sample is paired with observations
in the other sample. A paired t-test is used to compare two population
means when the differences between pairs are normally distributed. In
our case the population data does not have a normal distribution. What
is more, distribution of differences between pairs is severely
non-normally distributed (the differences in yield for stocks are
substantial). In such cases, Wilcoxon signed-rank test is used. The null
hypothesis for this test is that the medians of two samples are equal
(e.g.~Buy \& Hold vs.~W2V). We determined the statistical significance
with the help of z-score, that is calculated based on the Equation
\eqref{eq:wilcoxon-zscore}:

\begin{equation}
z \approx \frac{W - \frac{N (N + 1)}{4}}{\sqrt{\frac{N (N + 1) (2N + 1)}{24}}} = \frac{W - 637.5}{103.591}
\label{eq:wilcoxon-zscore}
\end{equation}

where \(N = 50\) denotes sample size (number of stocks) and \(W\)
denotes test statistics. Test statistics \(W = \min(W^{-}, W^{+})\),
where \(W^{-}\) denotes sum of the ranks of the negative differences and
\(W^{+}\) sum of the ranks of the positive differences (how many times
the yield of the first method is higher than the yield of the second
method). For the calculated z-value we look up in the normal probability
table (z-table) for the corresponding p-value. We accept our hypothesis
for p-values which are less than \(0.05\).

\begin{table}[H]

\caption{\label{tab:avg-yield-russel-10k-individual}Individuals yields of forecasting models on stocks of the Russell Top 50 index at an initial investment of \$10,000 in the test phase}
\centering
\begin{tabular}{lrrrr}
\toprule
 & Buy \& Hold & MA(50,100) & MACD & \textbf{W2V}\\
\midrule
VZ & -\$3.039,61 & \$647,66 & -\$2.990,29 & \textbf{\$3.080,84}\\
T & -\$359,55 & -\$42,32 & -\$1.280,35 & \textbf{\$8.273,56}\\
UNH & -\$3.769,27 & -\$288,51 & -\$3.398,75 & \textbf{\$5.050,80}\\
AMGN & -\$291,37 & \$252,85 & \$3.227,64 & \textbf{\$7.668,43}\\
GE & -\$3.927,61 & -\$2.472,75 & -\$3.128,26 & \textbf{\$2.956,75}\\
\addlinespace
CELG & \$25.708,81 & \$5.608,72 & \$12.781,08 & \textbf{\$50.882,06}\\
CMCSA & -\$542,23 & \$982,66 & -\$2.407,12 & \textbf{\$2.889,15}\\
KO & \$2.519,05 & \$3.778,97 & -\$1.866,26 & \textbf{\$5.813,29}\\
MCD & \$8.955,17 & \$1.114,14 & \$898,75 & \textbf{\$12.868,28}\\
AGN & \$3.441,65 & \$2.536,50 & \$747,18 & \textbf{\$5.693,08}\\
\addlinespace
QCOM & -\$772,94 & \$1.228,42 & -\$3.761,31 & \textbf{\$9.237,34}\\
SLB & \textbf{\$11.208,42} & \$8.533,32 & \$596,46 & \$3.357,58\\
HD & -\$4.491,7 & -\$5.002,76 & -\$4.843,12 & \textbf{\$2.520,49}\\
BAC & -\$4.641,57 & -\$4.017,49 & -\$2.588,89 & \textbf{\$5.880,18}\\
PFE & -\$3.913,79 & -\$3.236,17 & -\$4.358,06 & \textbf{\$3.252,18}\\
\addlinespace
WFC & \$309,97 & -\$3.488,26 & -\$6.321,32 & \textbf{\$6.627,38}\\
CVX & \$1.686,49 & -\$903,09 & \$5.414,68 & \textbf{\$9.172,12}\\
UTX & \$1.132,83 & \$301,21 & -\$1.690,97 & \textbf{\$5.169,85}\\
MDT & \$1.543,71 & \$81,96 & -\$3.058,66 & \textbf{\$5.608,06}\\
HON & -\$138,26 & -\$366,97 & \$147,02 & \textbf{\$6.283,95}\\
\addlinespace
BMY & -\$1.934,93 & -\$1.997,37 & -\$2.820,40 & \textbf{\$5.111,98}\\
BA & \$14,22 & \$2.171,52 & -\$1.990,97 & \textbf{\$6.615,65}\\
IBM & \$702,70 & -\$593,99 & -\$367,11 & \textbf{\$5.488,49}\\
WMT & \$410,02 & -\$2.501,45 & -\$1.724,35 & \textbf{\$6.359,24}\\
AAPL & \$31.114,37 & \$17.420,48 & \$26.956,85 & \textbf{\$97.776,72}\\
\addlinespace
MSFT & -\$1.231,96 & -\$4.080,51 & -\$674,72 & \textbf{\$8.067,93}\\
BRKB & \$4.642,86 & \$2.359,40 & -\$2.075,03 & \textbf{\$8.411,10}\\
MA & \$12.182,20 & \$9.519,22 & \$2.831,36 & \textbf{\$21.774,12}\\
DIS & \$674,97 & \textbf{\$2.110,20} & -\$818,91 & \$1.479,87\\
V & \$4.439,16 & \$4.203,69 & -\$2.030,36 & \textbf{\$8.327,75}\\
\addlinespace
MMM & -\$2.045,36 & -\$3.388,90 & -\$2.325,36 & \textbf{\$2.073,84}\\
PM & \$8.102,11 & \$3.265,02 & \$2.208,59 & \textbf{\$13.333,25}\\
INTC & -\$1.856,66 & -\$1.277,88 & -\$830,56 & \textbf{\$4.716,03}\\
CSCO & \$74,87 & -\$1.609,55 & -\$3.946,78 & \textbf{\$8.631,76}\\
PG & \$2.500,00 & -\$991,19 & -\$934,31 & \textbf{\$8.266,96}\\
\addlinespace
GOOGL & -\$1.031,18 & -\$230,00 & \$2.288,33 & \textbf{\$10.551,24}\\
UNP & \$9.815,55 & \$1.029,44 & \$2.830,19 & \textbf{\$18.177,28}\\
JNJ & \textbf{\$1.038,69} & -\$641,97 & \$20,70 & \$0,00\\
MRK & -\$278,32 & \$759,97 & -\$2.167,89 & \textbf{\$13.128,04}\\
XOM & \$5.746,22 & \$2.277,44 & -\$771,99 & \textbf{\$6.934,31}\\
\addlinespace
MO & -\$6.182,24 & \$616,95 & -\$8.350,84 & \textbf{\$2.346,59}\\
AMZN & \$6.610,64 & \$5.827,54 & \$6.074,77 & \textbf{\$35.389,87}\\
ABBV & \$2.900,97 & \$3.322,96 & \$1.512,46 & \textbf{\$5.227,56}\\
GILD & \$11.832,06 & \$8.258,70 & -\$5.391,78 & \textbf{\$22.280,68}\\
ORCL & \$3.631,19 & \$1.747,77 & -\$1.572,96 & \textbf{\$8.642,77}\\
\addlinespace
FB & \$18.046,26 & \$4.560,36 & \$2.071,40 & \textbf{\$23.657,48}\\
C & -\$6.524,43 & -\$2.595 & -\$5.441,80 & \textbf{\$8.239,60}\\
CVS & \$3.344,02 & \$833,63 & -\$3.662,15 & \textbf{\$13.220,15}\\
PEP & \$3.235,77 & \$1.406,87 & -\$159,01 & \textbf{\$6.692,48}\\
JPM & \$352,82 & -\$3.378,22 & -\$4.968,51 & \textbf{\$12.836,16}\\
\bottomrule
\end{tabular}
\end{table}

Table \ref{tab:wilcoxon-results} shows values for test statistics \(W\)
and corresponding p-values. If we focus on the test phase, obtained
p-values are much smaller than \(0.05\) for all three popular models.
This means that there is a statistically significant difference between
the resulting yields achieved by our proposed model and existing three
models. For the test phase we can conclude with a high level of
confidence that appropriately parametrised proposed model W2V performed
better than existing three models. As mentioned earlier, the proposed
method achieved worse results in the validation phase. In the validation
phase the difference in returns between the proposed model and reference
models was statistically significant only for MACD and MA. Comparing to
MACD, we obtained p-value smaller \(0.0001\), which means that our model
yields better results. Similar results are obtained when comparing W2V
to MA(50, 100).

\begin{table}[H]

\caption{\label{tab:wilcoxon-results}The Wilcoxon Signed Rank Test for forecast models}
\centering
\begin{tabular}{lrrrr}
\toprule
\multicolumn{1}{c}{ } & \multicolumn{2}{c}{test phase} & \multicolumn{2}{c}{validation phase} \\
\cmidrule(l{3pt}r{3pt}){2-3} \cmidrule(l{3pt}r{3pt}){4-5}
 & W & p-value & W & p-value\\
\midrule
Buy \& Hold & 2 & < 0,0001 &  & \\
MA(50, 100) & 1 & < 0,0001 & 427 & 0.021\\
MACD & 1 & < 0,0001 & 155 & < 0,0001\\
\bottomrule
\end{tabular}
\end{table}

The p-value was \(0.021\), which is more than half that of the limit of
\(p = 0.05\), where the hypothesis could not be confirmed. When compared
to Buy \& Hold, the W2V method yields lower returns. That can be seen
already from the average yields in Table \ref{tab:avg-yield-russel-10k}.
To prove that Buy \& Hold gives statistically better results compared to
W2V method, we performed additional Wilcoxon Signed Rank t-test, and
obtained a p-value of less than \(0.0001\).

Given the presented results, we can conclude that if the model
parameters are well set, presented forecast model gives better results
than presented popular models. However, the parameters that perform well
in the test phase may not perform equally well in the later validation
phase.

\section{Conclusion and Future Work}\label{conclusion-and-future-work}

Stock trend forecasting is a challenging task and has become an
attractive topic during the last few decades. In this paper, we present
a novel approach for stock trend prediction. Besides focusing on
prediction accuracy, the presented approach was also tested for
financial success. In the test phase, we used three sample stocks --
Apple (AAPL), Coca-Cola (KO) and Microsoft (MSFT) -- that satisfied
conditions for a good test case (the stock trend diverse in observed
period, known company's business model, enough data available). The
confirmation analysis was performed with analysis on Russell Top 50
Index.

We realized that even if the forecasting model has high prediction
accuracy, it can still achieve bogus financial yields, if poor trading
strategy is used. However, despite the simplicity of the proposed
model's trading strategy, its performance was very good with a
statistical significance.

In the test phase, the proposed model performed well for all three
sample stocks. The yields were higher than the yields of the comparative
models, i.e., Buy \& Hold, MA and MACD. In the validation phase, the
proposed model outperformed MA and MACD models, while the Buy \& Hold
turned out to be statistically most profitable. In more extensive
testing on Russell Top 50 Index, the proposed method was outperformed
only by Buy \& Hold, while achieved statistically better average yields
than MA and MACD.

A more detailed analysis of trading graphs and statistical analysis
showed that the proposed model has a great potential for practical use.
However, it is too early to conclude that the proposed model provides a
financial gain, as we have shown that selected model parameters are not
equally appropriate for different time periods in terms of yield. We
have also shown that the forecast model is strongly influenced by the
training data set. If the model is trained with data that contains bear
trend, the predictive model might be very cautious despite the general
growth trend of validation data set. The problem is due to over-fitting,
so training with more data would help. Some of the state-of-art machine
learning algorithms like \textbf{Word2Vec} are dependent on a
large-scale data set to become more efficient and eliminate the risk of
over-fitting.

We hope that the proposed approach can offer some beneficial
contributions to a stock trend prediction and can serve as a motivation
for further research. In the future, we would like to improve method's
trading strategy and incorporate the stop loss function and some other
proven, often used technical indicators. In the training phase, we could
include OHLC data of other stocks to acquire more diverse patterns,
reducing unknown ones. That would help algorithms to identify the
underlying future chart pattern better. To improve classification
accuracy and logarithmic loss, the SoftMax regression could be replaced
with advanced machine learning classification algorithms. It is also
worth exploring, how candlestick data with different time grains affect
prediction accuracy. This way we could compare daily, weekly or monthly
trend forecasts.

\end{document}